\documentclass[aps,prl,preprint,a4paper]{revtex4-1}

\usepackage{graphicx}
\usepackage{amsfonts,amsmath,amssymb}
\usepackage{natbib}
\usepackage[english]{babel}

\begin{document}
\title{Plasmonic nanoantenna design and fabrication based on evolutionary optimization}
\author{Thorsten Feichtner}
\altaffiliation[Now at: ]{Max-Plank-Institute for the Science of light, G\"unther-Scharowsky-Str. 1; Bldg. 24; D-91058 Erlangen}
\affiliation{Nano-Optics \& Biophotonics Group, Department of Experimental Physics 5, R\"ontgen
Research Center for Complex Material Research (RCCM), Physics Institute,
University of W\"urzburg, Am Hubland, D-97074 W\"urzburg, Germany}

\author{Oleg Selig}
\altaffiliation[Now at: ]{FOM Institute AMOLF, Biomolecular Photonics Group, Science Park 104, 1098 XG Amsterdam, The Netherlands}
\affiliation{Nano-Optics \& Biophotonics Group, Department of Experimental Physics 5, R\"ontgen
Research Center for Complex Material Research (RCCM), Physics Institute,
University of W\"urzburg, Am Hubland, D-97074 W\"urzburg, Germany}

\author{Bert Hecht}
\email{bert.hecht@physik.uni-wuerzburg.de}
\affiliation{Nano-Optics \& Biophotonics Group, Department of Experimental Physics 5, R\"ontgen
Research Center for Complex Material Research (RCCM), Physics Institute,
University of W\"urzburg, Am Hubland, D-97074 W\"urzburg, Germany}

\begin{abstract}
Nanoantennas for light enhance light-matter interaction at the nanoscale making them useful in optical communication, sensing, and spectroscopy. So far nanoantenna engineering has been largely based on rules derived from the radio frequency domain which neglect the inertia of free metal electrons at optical frequencies causing phenomena such as complete field penetration, ohmic losses and plasmon resonances. Here we introduce a general and scalable evolutionary algorithm that accounts for topological constrains of the fabrication method and therefore yields unexpected nanoantenna designs exhibiting strong light localization and enhancement which can directly be "printed" by focused-ion beam milling. The fitness ranking in a hierarchy of such antennas is validated experimentally by two-photon photoluminescence. Analysis of the best antennas' operation principle shows that it deviates fundamentally from that of classical radio wave-inspired designs. Our work sets the stage for a widespread application of evolutionary optimization to a wide range of problems in nano photonics.
\end{abstract}

\maketitle

The reception of radio frequency signals using antenna structures is based on the detection of electron oscillations in metal bodies, e.g. wires, driven by external time-dependent electric fields. Vice versa, accelerated electrons in metal bodies can cause very efficient emission of electromagnetic waves\cite{Balanis1992}. Similar ideas can be applied to realize antennas for light\cite{Novotny2011,Biagioni2012}, although for such very high frequencies many established concepts of antenna theory need to be adapted. For example the occurrence of plasmon resonances\cite{Maier2007} and volume currents\cite{Dorfmuller2010} have to be considered, and new phenomena become important, e.g. the kinetic inductance\cite{Zhou2005} and Ohmic loss\cite{Zhou2014}. To rescue the concepts of radio frequency antenna technology into the optical realm the principle of effective wavelength scaling was introduced\cite{Novotny2007} and applied to e.g. Yagi-Uda antennas\cite{Curto2010}. However, this approach does not always lead to the best possible antenna performance and, as we have shown before, unexpected designs found by evolutionary optimization can perform much better\cite{Feichtner2012}.
    
So far, evolutionary optimization has been applied to plasmonic systems that searched a limited parameter space, e.g. arrays of scattering discs\cite{Forestiere2010, Forestiere2012} as well as specifically shaped single particles\cite{Ginzburg2011}. Here, we show how to perform evolutionary optimization of plasmonic geometries in a by far larger parameter space while taking into account topological constrains of the fabrication method. This is achieved by using smallest primitive elements ("pixels") and a set of rules that take into account the topology of pixel arrangements to incorporate characteristics of the focused-ion-beam milling process. This approach guarantees that every structure that is obtained during evolutionary optimization can be readily fabricated in full detail by focused ion beam (FIB) milling of a thin layer of monocrystalline gold. This allows us to experimentally investigate a hierarchy of antenna structures obtained by genetic optimization and validate their relative performance experimentally using two-photon photoluminescence. A detailed analysis of finite-difference-time-domain simulations reveals an unique operation principle of the fittest antenna.

\section{Results}

\subsection{Evolutionary Algorithm}

An evolutionary algorithm (EA) is an iterative numerical technique inspired by biology to find optimized solutions of non-analytical problems\cite{Sivanandam2008}. It is based on inheritance of favorable properties of a valid solution (individual), which is fully described by a string of characters (genome). One iteration step consists of two parts: i) A set of individuals (generation) is evaluated with respect to its capability to solve the posed problem. The evaluation is realized by a fitness function which assigns a real number (fitness parameter) to each individual thereby creating a ranking of solutions within a generation. ii) The genome of the fittest individuals of a generation (parents) is used as source for the next generation, assembling new genomes via small random changes of parent genomes (mutation) or the combination of two parent genomes to generate a child genome (crossing).

Here we adapted our EA based on finite difference time domain (FDTD) simulations combined with a Matlab code\cite{Feichtner2012} to find planar optical antennas for which the near-field localization and intensity enhancement is optimized at $\mathbf{r}_0$, the very center of the antenna. Accordingly, the fitness parameter was chosen to be the near field intensity enhancement (NFIE) polarized along the $x$-direction $I_x(\mathbf{r}_0)=E_x(\mathbf{r}_0)^2$, normalized to the illumination intensity without antenna.

\begin{figure}[ht]%
\centering
\includegraphics{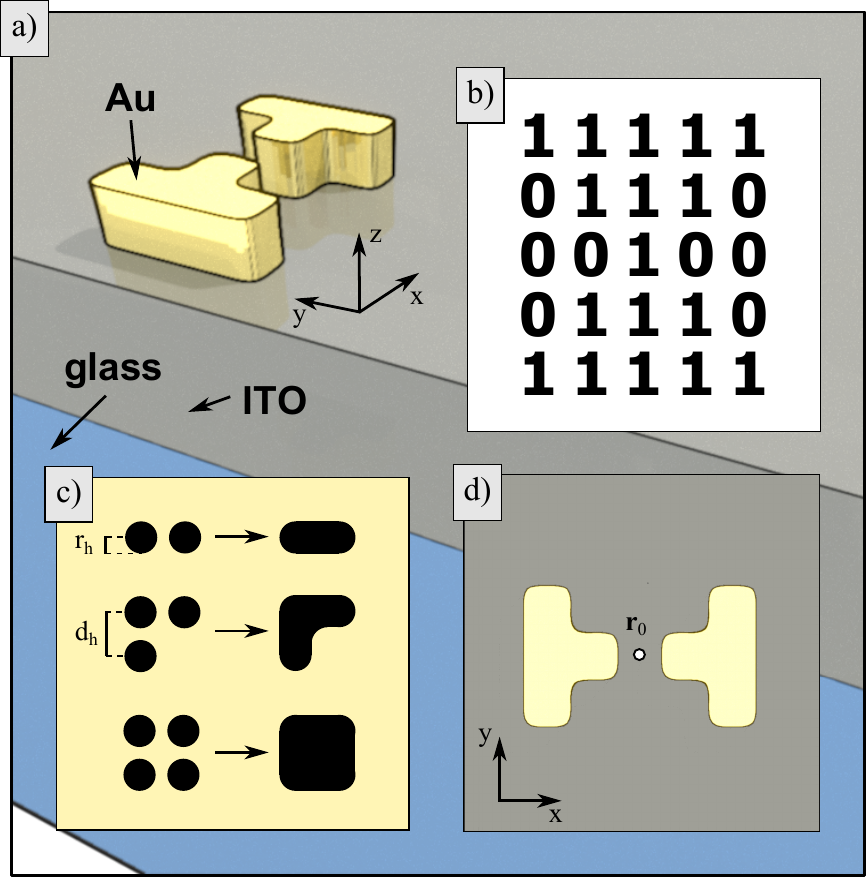}
\caption{\label{fig:matrix_to_geometry} Genome and topological constraints. (a) 3D sketch of an exemplary geometry on top of an ITO layer covering a glass substrate. The rounded features occur due to FIB milling. (b) binary 5$\times$5 genome of a small optical bow-tie-like antenna, where '1' denotes the positions at which the gold will be removed. (c) Rules for replacing neighboring hole arrangements by structures that can be fabricated by FIB. $r_h$: hole radius, $d_h$: center-to-center hole distance. (d) top view of the structure resulting from the genome in (b) after applying the rules sketched in (c). $r_0$ denotes the point of optimization for the NFIE.}%
\end{figure}

Figure \ref{fig:matrix_to_geometry} illustrates the encoding and geometrical interpretation of the planar antenna structures used within the genetic algorithm. Antennas are assumed to consist of 30 nm thick gold on top of 200 nm indium tin oxide (ITO) on top of a glass substrate, corresponding to later experimental conditions (see Fig.~\ref{fig:matrix_to_geometry}(a)). The high transparency and the good conductivity of the substrate are ideal for ion beam milling, SEM imaging, as well as optical characterization. The genome is a two dimensional square matrix with binary entries (see Fig.~\ref{fig:matrix_to_geometry}(b)). Each '1' denotes a cylindrical hole which approximates the structural primitive of FIB fabrication. As the antenna center is meant to be the area of maximum NFIE, the matrix center is always set to '1'. Figure~\ref{fig:matrix_to_geometry}(c) illustrates, how the topology of possible hole arrangements is converted to a realistic geometry that can be fabricated by FIB milling: adjacent holes are connected, leading to geometries as the one sketched in Fig.~\ref{fig:matrix_to_geometry}(d). A test pattern was devised (see Supplementary Fig.~1) to identify the parameters for the hole radius $r_h$ and center-to-center hole distance $d_h$ which result in reproducible patterns when FIB milling a 30 nm thick single crystalline gold flake\cite{Huang2010}.

The evolutionary algorithm was run using a 11$\times$11 square array with $r_h=11$ nm and $d_h = 30$ nm. In the simulations the antenna is excited using a Gaussian focus (NA = 1.4; $\lambda_\text{exc}=830$ nm) at normal incidence, centered onto the structure. The resulting overall antenna area of 330 nm$\times$330 nm fits the FWHM of the Gaussian focus of $\approx 390$ nm.

\begin{figure*}[ht]
\centering
\includegraphics[width=\textwidth]{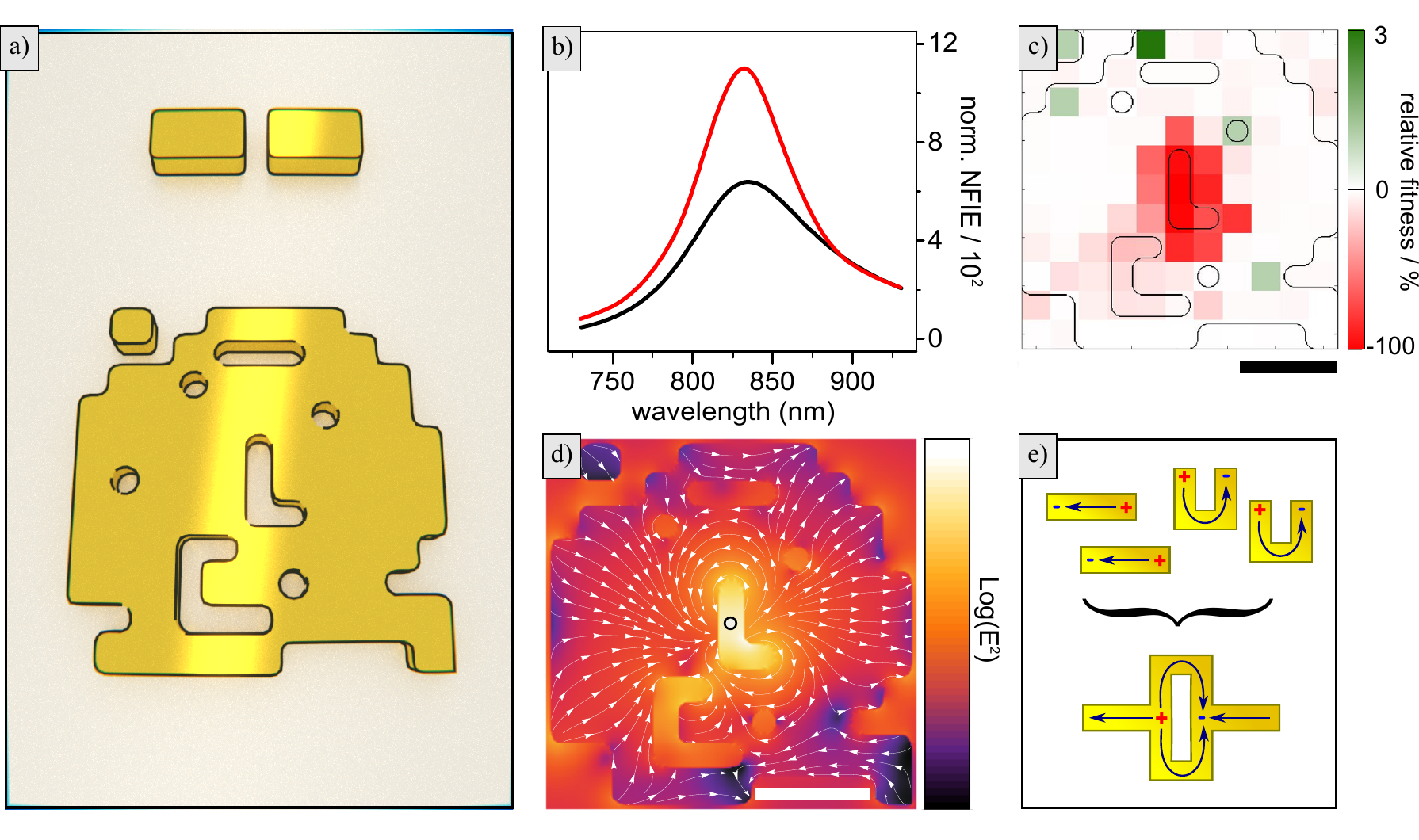}
	\caption{Properties of the fittest evolutionary antenna (FEA): (a) 3D sketch of the reference dipolar antenna (top) and the FEA (bottom) on substrate. (b) Near field intensity enhancement (NFIE) spectrum at the very center of the reference antenna (black) and the FEA (red). (c) Toggle plot of FEA (for details see text). (d) Near field intensity and current direction (white arrows) in the center plane of the FEA. (e) Model for the operation principle of the FEA constructively combining current patterns from two split-ring resonators and two rod antennas. All scale bars are 100 nm.}
	\label{fig:expEA_result}
\end{figure*}

The best geometry originating from the EA (exact parameters in the methods section; see also Supplementary Fig.~2 and 3), the fittest evolutionary antenna (FEA), is a compact asymmetric structure with a high filling factor of 65 \% with an L-shaped gap in the very center (see lower part of Fig.~\ref{fig:expEA_result}(a)). We compare its fitness to a dipolar antenna with corresponding geometrical constrains (depicted in Fig.~\ref{fig:expEA_result}a, top), its arm dimensions being $112\times38\times30$ nm$^3$ with a gap of 22 nm optimized via its arm length to exhibit a near field intensity enhancement in the gap at $\lambda = 830$ nm.

The NFIE spectra of both antennas show a Lorentzian-shaped resonance with a maximum of 1100 at $\lambda_\text{FEA} = 832$ nm for the FEA and of 640 at $\lambda_\text{res} = 835$ nm for the reference antenna (see Fig.~\ref{fig:expEA_result}(b) with a full width at half maximum of $\Delta\lambda_\text{FEA}= 74$ nm compared to $\Delta\lambda_\text{ref}= 105$ nm. This corresponds to a Q-factor $Q_\text{FEA} = \lambda_\text{FEA}/\Delta\lambda_\text{FEA} = 11.2$ for the FEA and $Q_\text{ref} = \lambda_\text{ref}/\Delta\lambda_\text{ref} = 8.0$ for the reference antenna, respectively. As a higher Q-factor originates from lower combined radiative and ohmic losses\cite{Biagioni2012}, the far-field coupling of the FEA is reduced in comparison to the reference antenna. Nevertheless, the FEA exhibits an 1.7-fold higher near-field intensity enhancement in the antenna center, which further highlights its improved energy concentration mechanism.

The asymmetry of the FEA geometry originates from the still too coarse discretization provided by the antennas primitive elements: changing antenna dimensions in steps of 30 nm results in large shifts of resonance frequency making it difficult to perfectly match the optimization frequency with a symmetric geometry. The importance of individual FEA building blocks for the obtained NFIE can be assessed using a toggle plot (Fig.~\ref{fig:expEA_result}(c), which visualizes the relative fitness change of 120 antenna structures with a single pixel switched from '0' $\to$ '1' or vice versa with respect to the FEA. The toggle plot clearly demonstrates that the pixels close to the center play the most important role for the antenna performance. Some green areas indicate that the FEA structure does not correspond to the absolute global maximum in the configuration space of the EA and a slight increase of near field intensity enhancement up to 3 \% can be achieved by switching single elements in the periphery (see also Supplementary Fig.~2 and 3).

To understand the working principle of the FEA we analyze the NFIE distribution (color coded in Fig.~\ref{fig:expEA_result}(d) and a temporal snapshot of the current pattern (white lines and arrows). The near field intensity at the optimization point (marked with circle) yields a 1100-fold normalized enhancement which is nearly fully $x$-polarized. The current pattern suggests an accumulation of charges at positions close to the optimization point within the antenna even in the absence of tip-like structures. Only the kink of the L-shaped central void contributes to the enhancement by means of the lightning rod effect, similar as in bow-tie antenna geometries\cite{Fernandez-Garcia2014}.

The FEA current pattern can be described by a constructive superposition of two fundamental modes\cite{Feichtner2012} (see Fig.~\ref{fig:expEA_result}(e)): (i) dipolar antenna currents comparable to the lowest order bonding-mode current pattern of linear two-wire antennas\cite{Biagioni2012} with the well-known benefits of good far-field coupling as well as accumulation of opposite charges at either side of the gap as well as; (ii) the current pattern of fundamental split-ring like modes\cite{Rockstuhl2006} above and below the gap, leading to additional charge accumulations at the center and resulting in a larger NFIE as for a plain dipolar antenna. A similar concept has been used to numerically optimize tips for scanning near-field optical microscopy\cite{Garcia2012}.

\subsection{Experiment}

\begin{figure}[ht]
\centering
\includegraphics{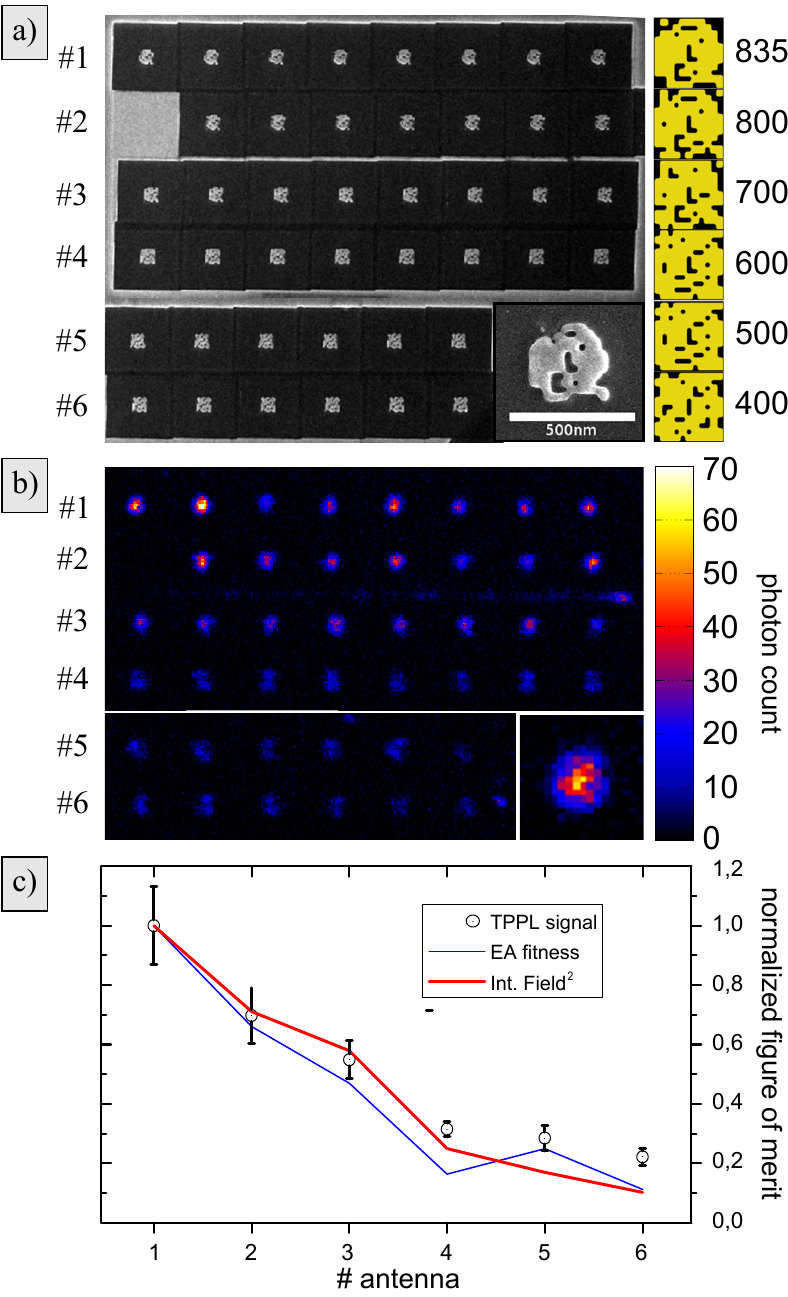}
	\caption{Experimental realization of evolutionary antennas: (a) Exemplary SEM image of the EA antenna array realized by means of FIB milling together with geometry sketches and simulated fitness. Each row contains six to eight copies of the same geometry, the fitness decreasing from top to bottom. (b) TPPL map of the fabricated array. The insets show detailed maps of one of the very best antennas. (c) TPPL data (black dots) alongside with simulated fitness (near field intensity enhancement - red line) and the simulated TPPL (equation \eqref{eq:TPPL_integral} - blue line).}
	\label{fig:exp_data}
\end{figure}

The FEA as well as five antenna structures from earlier generations showing a decreasing fitness were fabricated by means of focused ion beam milling of a single-crystalline gold flake\cite{Huang2010} (see Fig.~\ref{fig:exp_data}(a) and methods for details). To evaluate the reproducibility, each structure was fabricated six to eight times in a row, the first row denoted \#1 containing the FEA.

The thickness of the structures after focused ion beam milling was measured by AFM to be $28\pm 2$ nm closely matching the intended 30 nm (see Supplementary Fig.~4). To account for the small deviation in thickness all geometries were re-simulated with decreased thickness (see Supplementary Fig.~5) of 28 nm resulting in a slightly reduced fitness for all geometries, while preserving the hierarchy of relative fitnesses, except for a slight increase for structure~\#5.

Two-photon-photo-lumi\-nescence (TPPL) is used to experimentally probe the relative change of near-field intensity enhancement of the fabricated antenna structures\cite{Huang2010a}. In good approximation the intensity $I_\text{TPPL}$ of the TPPL signal is proportional to the integral of the forth power of the electrical field $\mathbf{E}$ within the volume of the antenna $V_\text{ant}$\cite{Imura2005}:
\begin{equation}
I_\text{TPPL} \propto \int_V \mathbf{E}^4 \, dV_\text{ant}\quad .
\label{eq:TPPL_integral}
\end{equation}
Equation \eqref{eq:TPPL_integral} allows to numerically calculate the relative changes of emitted TPPL signal based on simulated antenna modes\cite{Huang2010a}.

Additionally, TPPL is dipole forbidden and requires strong field gradients introduced by the curvature of the nano structures, leading to vanishing signal at flat metal films. The vast majority of the signal therefore originates from the volume near the antenna hot spots, where the field gradients are largest. The fabricated EA antennas show nearly identical geometries around the central hot spot. They can therefore be compared reliably by means of TPPL, which provides a measure for the relative NFIE, assuming $I_\text{TPPL} \propto I_x^2(\mathbf{r}_0)$. The gap of the reference antenna strongly deviates in shape from the EA antennas and their TPPL signals cannot be compared quantitatively.

TPPL of the fabricated antennas is recorded by scanning confocal microscopy using a 300 fs Ti:Saph laser at a center wavelength of 830 nm (see methods for details). Figure~\ref{fig:exp_data}(b) shows the resulting TPPL map. The antennas are clearly distinguishable and the trend of decreasing signal for antennas with lower fitness is obvious.

Figure \ref{fig:exp_data}c shows the comparison between experimental and numerical results. The experimental data points are acquired by integrating the TPPL count rates over the area of the individual antennas. Mean and standard deviation are calculated and normalized to the value of the FEA. As numerical results both the simulated fitness parameter as well as the simulated TPPL signal (equation~\eqref{eq:TPPL_integral}) for 28 nm thick antennas are plotted, each also normalized to the respective value of the FEA. There is good agreement for the relative changes of measured TPPL signal, simulated fitness and simulated TPPL signal. The trend of increasing error bars with increasing fitness can be explained by the influence of different hole sizes due to fabrication inaccuracies. The NFIE of the antennas depends heavily on the central area geometry due to the capacitive coupling across the width of the center gap (see Supplementary Fig.~6).

For the two antennas with lowest fitness the measured normalized count rates are consistently higher than the simulated relative signal strengths. This maybe due to different effects: i) the simulated broadband spectra (see Supplementary Fig.~7) show an additional small resonance peak at about $ \lambda =$ 670 nm for those two geometries, which will lead to an enhancement of the TPPL signal compared to the other antennas\cite{Wissert2010}. ii) antennas~\#5 and \#6 do not show a single central peak in the TPPL map. This most probably originates in the existence of multiple or higher order modes\cite{Huang2010a}, not included in the simulations with the excitation focus being fixed in the very center of their geometries. Both effects were not considered.

\section{Conclusions}

We established an evolutionary algorithm describing realistic planar optical antenna geometries with feature sizes of $\approx$ 22 nm that can be directly printed via FIB milling. The fittest antenna resulting from an optimization of near field intensity enhancement is a rather compact, yet complex geometry which exhibits a surprisingly clean Lorentian resonance. The responsible mode can be described by a superposition of a dipolar antenna resonance and split ring resonances. Comparison of experimental two-photon photo luminescence data and corresponding numerical simulations show good agreement and prove the possibility to establish a direct link between evolutionary optimization and fabrication of optimized structures which indeed display the expected high performance.

\section{Methods}

\paragraph{FDTD simulations} Commercial software (FDTD Solutions, Lumerical) was used to numerically solve Maxwell's equations by means of the finite-difference-time domain algorithm. For the dielectric function of gold an analytical model was used \cite{Etchegoin2006,Etchegoin2007}, while the optical index for glass and ITO where set to $n=1.4$ and $n=2$ respectively. The fitness parameter $I_x$ was normalized to the maximum field intensity of the Gaussian excitation without antenna but in presence of the substrate.

\paragraph{Evolutionary Optimization} A MatLab script generates the binary matrices representing the individuals and converts them into rounded geometries within the FDTD solver according to topological rules (see Fig.~\ref{fig:matrix_to_geometry}). One generation of the evolutionary algorithm consists of 30 individuals and to obtain the best antenna structure a run of 60 generations has been performed (e.g. see Supplementary Fig.~3). The first generation consisted of random structures with a filling factor of 0.7. The antennas were ranked according to their fitness and the best eight structures were taken as parents of the subsequent generation (more details about the mechanism of the EA can be found in \cite{Feichtner2012} as well as in the supplementary material). Three methods were employed to create the next 30 individuals: mutation (creation of random structures), as well as linear and spiral genome crossing (see Supplementary Fig.~2 bottom). To optimize the performance of the algorithm, each simulation was terminated after 35 fs internal simulation time. At that moment typically $>98$\% of electromagnetic energy has already left the simulation volume and the obtained results represent a good approximation to a fully converged simulation. This is sufficient for comparing the fitnesses of individuals within one generation of the EA while reducing the simulation time of a single individual to $\approx$ 40 min instead of $\approx$ 2h for a full simulation. All data used for quantitative evaluation and in particular all data presented in this work are retrieved from fully converged simulations.

\paragraph{Sample substrate preparation} Microscope cover slips (Menzel, 24 $\times$ 24 mm$^2$, 0.17 nm thick) were covered with 200 nm of sputtered ITO. A gold marker structure was evaporated and processed by means of optical lithography. Wet-chemically grown gold flakes \cite{Huang2010} were drop-cast and the resulting sample was plasma cleaned for 60 s in a 30 W low pressure oxygen plasma.

\paragraph{AFM measurements} The thickness of flakes and structures was measured at ambient conditions using tapping mode AFM operating at a resonance frequency of 240 -- 280 kHz and a scanning rate of 0.2 Hz (DMLS scanning head, Nanoscope IIIa, Digital Instruments).

\paragraph{Antenna structuring} Focused ion beam (FIB) milling (Helios Nanolab 600, FEI) was used to polish gold flakes down to the desired thickness of 30 nm. A current of 48 pA at an acceleration voltage of 30 kV was used and the sputtering rate was calibrated with assistance of AFM measurements beforehand.

To fabricate the EA antennas an ion current of 9.7 pA at an acceleration voltage of 30 kV was used. A MatLab script translates the binary matrix of the EA into a script for the FIB pattern generator, defining a set of polygons, which can be directly written by the FIB machine. This pattern was written in four paths, each path with a different direction (top to bottom, left to right, ...) to ensure the milled features being regularly shaped. Milling in only one direction leads to undesired redeposition effects. The pattern generator showed a substantially decreased milling depth for single holes of the EA pattern. We automatically adjusted the milling depth for single holes by a factor of 1.4, which was determined by empirical tests. The MatLab script therefore represents a "printer driver" for the FIB milling instrument.

\paragraph{Two-photon photo luminescence microscopy (TPPL)} The antennas where scanned through the focus of a pulsed laser with a center wavelength of $\lambda =$ 830 nm (pulse length at sample position $\approx$ 900 fs, repetition rate 76 MHz, Coherent Inc. MIRA 900) focused via an oil immersion objective (Nikon, Plan APO 100x, NA=1.4). In the detection path, after a non-polarizing 50 :50 cube beam splitter, three filters (notch filter: OD $>$ 6 at 830nm, Kaiser Optical System; two short pass filters: SP785 and SP680, Semrock) ensured the blocking of the direct reflection of the excitation laser. The remaining TPPL signal was focused onto a single photon counting module (PDM Series, Micro Photon Devices). The full setup is sketched in Supplementary Fig.~8.

\begin{acknowledgments}
The authors wish to thank Johannes Kern and Swen Großmann for fruitful discussions. Financial support by the DFG is gratefully acknowledged (HE5648/1-1).
\end{acknowledgments}

\paragraph{Author contributions}
T.F.~conceived the idea and fabricated the nano antennas. O.S.~implemented the algorithm and performed the TPPL measurements. B.H.~supervised all activities. All authors contributed to discussions and to the writing of the manuscript.

\end{document}